\documentclass[twocolumn,english,prb,showpacs]{revtex4}
\usepackage[latin1]{inputenc}
\usepackage{amsmath}
\usepackage{amssymb}

\makeatletter
\usepackage{amssymb}

\usepackage{graphicx}
\usepackage{amssymb}
\usepackage{babel}
\makeatother
\begin{document}

\title{Magnetic field induced Coulomb blockade in small disordered delta-doped
heterostructures}

\author{V. Tripathi$^{1}$ and M. P. Kennett$^{2}$}

\affiliation{$^{1}$ Department of Theoretical Physics, Tata Institute of Fundamental
Research, Homi Bhabha Road, Mumbai 400005, India}

\affiliation{$^{2}$ Physics Department, Simon Fraser University, 8888 University
Drive, Burnaby, BC, V5A 1S6, Canada}

\date{\today{}}

\begin{abstract}
At low densities, electrons confined to two dimensions in a delta-doped
heterostructure can arrange themselves into self-consistent droplets
due to disorder and screening effects. We use this observation to
show that at low temperatures, there should be resistance oscillations
in low density two dimensional electron gases as a function of the
gate voltage, that are greatly enhanced in a magnetic field. These
oscillations are intrinsic to small samples and give way to variable
range hopping resistivity at low temperatures in larger samples. We
discuss recent experiments where similar physical effects have been
interpreted within a Wigner crystal or charge density wave picture. 
\end{abstract}

\pacs{73.20.-r, 05.60.-k, 73.23.Hk, 75.47.-m}

\maketitle

\section{Introduction}

The interplay between disorder and interactions in spatially inhomogeneous
electronic states is an important ingredient in the physics of many
strongly correlated electronic materials. Two dimensional electron
gases (2DEGs) provide an ideal laboratory to gain insight into the
effects of disorder and interactions as device properties such as
disorder or carrier concentration can be tuned in the growth process,
or by external means, such as a gate. Recent experiments \cite{Baenninger2}
on small, disordered, delta-doped devices suggest a new set of unusual
resistance oscillations in a perpendicular magnetic field. The resistance
as a function of gate voltage is featureless at zero magnetic field,
but at non-zero fields develops peaks that are evenly spaced in gate
voltage and whose magnitudes grow with increasing magnetic field whilst
the positions of the peaks are relatively unaffected. Similar effects
have been previously observed in low density 2DEGs, \cite{Ghosh2,Ghosh3,Ilani2}
and arrays of quantum dots.\cite{Dorn}

In earlier work on the same devices, \cite{Baenninger2,Ghosh3,Ghosh2}
there was the remarkable observation that the electron tunneling distance
(extracted from the magnetoresistance) is directly proportional to
the average electron separation, $r_{ee}$ in the 2DEG. This was used
to argue in favor of a charge density wave (CDW) or Wigner crystal
(WC) picture.\cite{Baenninger3} The value of $r_{s}$ in these devices
is around 5, which is much less than the prediction of $r_{s}=37$
for Wigner crystallization in two dimensions in a clean system, but
close to the prediction of $r_{s}\simeq7.5$ in disordered systems.
\cite{Tanatar}

A number of theoretical studies have shown that the charge distribution
in disordered delta-doped heterostructures at $r_{s}>1$ but too small
for a WC is likely to be neither Fermi liquid nor WC, but a droplet
\cite{Suris,droplets,VM1,hall-droplets} or {}``emulsion'' phase.\cite{kivelson}
The simplest picture for the droplet phase is one where nonlinear screening by
 electrons is unable to dominate the disorder-induced
potential barriers between regions of localized electrons.\cite{Suris,droplets,VM1}
A droplet-like phase may also arise at comparatively higher electron
densities in disordered quantum Hall insulators \cite{Ilani2,hall-droplets}
due to the interplay of the localization effects of disorder and screening,
where the screening is now sensitive to whether or not the electrons
in different parts of the system belong to a partially or completely
filled Landau level. A third, and somewhat different, picture proposed
for low disorder is one of an {}``emulsion'' or stripe-like phase
with crystalline regions in an electron liquid background.\cite{kivelson}

Nanoscale electronic inhomogeneity has also been observed in diverse
strongly correlated electron systems.\cite{Davis} Some of the above
mechanisms might be responsible for phase separation in many of these
systems. In particular, electronic inhomogeneity in the superconductor
BSCCO has been attributed to localization effects of the disorder
in oxygen doping.\cite{Mcelroy}

In this paper, we use the first of the above mentioned droplet pictures
to argue that the experiments in Refs.~\onlinecite{Baenninger2,Ghosh3,Ghosh2}
manifest Coulomb blockade effects greatly enhanced by a magnetic field.
We ignore quantum Hall physics in our treatment of the droplet phase
taking note of the fact that compared to quantum Hall insulators believed
to have a droplet phase, these experiments were performed on devices
with strong disorder and low electron density, and no quantum Hall
effect was seen at high fields. We make specific predictions about
the conditions under which this novel magnetic field induced Coulomb
blockade will occur, and find that our model is very successful in
explaining resistance versus temperature data in Ref.~\onlinecite{Baenninger2}.
In particular, such Coulomb blockade should be a signature of an electron
droplet phase. 

In previous work \cite{VM1} we derived expressions for the physical
parameters of electron droplets in the non-linear screening regime
that is relevant to the experiments of interest here, and applied
this picture to explain the experimentally observed density dependence
of the tunneling distance without invoking a CDW picture.

We propose here that the resistance oscillations arise from the decrease
of inter-droplet tunneling due to shrinking of the localization length
in strong magnetic fields. The decrease in inter-droplet conductance
is sufficient at larger magnetic fields to lead to a visible Coulomb
blockade effect in samples where this is not clearly resolved at zero
magnetic field. Considered together with our earlier explanation \cite{VM1}
for the density dependence of the tunneling length, we believe our
simple picture provides a complete description of the experiments
in Ref.~\onlinecite{Baenninger2} without invoking ordered electronic
states. The experiments we study fall in a parameter regime where
WC ordering is not expected theoretically, hence more convincing evidence
for WC ordering in these experiments is needed than has been offered
to date. Our work does not preclude the possibility of WC ordering
in heterostructures with low density and high mobility that have higher
values of $r_{s}$.

The remainder of the paper is organized as follows. In Sec. \ref{sec:Model-and-experimental}
we briefly describe the model and experimental parameters. Sec. \ref{sec:Magnetic-field-dependence}
recapitulates the magnetic field and density dependences of the inter-droplet
tunneling conductance obtained in earlier work. The main analysis
of the paper, namely, magnetic field induced Coulomb blockade is presented
in Sec. \ref{sec:Coulomb-blockade} and in Sec. \ref{sec:Discussion}
we give a a discussion of the results.

\section{Model and experimental parameters \label{sec:Model-and-experimental}}

Unless otherwise specified, we assume the following device parameters:
the $\delta-$doping density is $n_{d}=10^{12}\,\textrm{cm}^{-2}$,
the 2D electron density is $n_{e}\sim10^{11}\,\textrm{cm}^{-2}$,
$\lambda=50\,{\rm nm}$ is the distance of the $\delta$ layer from
the 2DEG and $d=300\,\textrm{nm}$ is the distance of the metallic
gate electrode from the 2DEG.

The important parameters describing the spatial extent of the electron
droplets are: \cite{VM1} $R_{c}=n_{d}^{1/2}/(\pi^{1/2}n_{e}),$ the
lengthscale on which potential fluctuations are screened by electrons;
and $R_{p}=\sqrt{a_{B}(\lambda+z_{0})}\sim30\,\textrm{nm}$ the mean
droplet radius, where $z_{0}\sim50$ nm is the extent of the wavefunction
in GaAs perpendicular to the surface. This is to be contrasted with
earlier predictions that $R_{p}\sim\lambda$. \cite{Suris} Increasing
the electron density has little effect on the droplet size. Extra
electrons are accommodated by increasing the density of droplets.
\cite{Suris,VM1} The separation between droplet centers, $l_{ip}=2(n_{d}^{1/2}R_{p}/\pi^{1/2}n_{e})^{1/2}=2\sqrt{R_{c}R_{p}},$
decreases with increasing $n_{e},$ and $l_{ip}\simeq85\,\textrm{nm}.$
The droplets merge at high enough $n_{e}$ when $R_{p}>l_{ip}$.

The important energy scales for the droplets are: $E_{\textrm{barrier}}\simeq58$
K, the difference between the binding energy $E_{B}$, and the highest
occupied energy level $\Delta=\hbar^{2}\sqrt{\pi n_{d}}/2mR_{p}\simeq36$
K. The typical number of electrons in a droplet is $N_{e}=\sqrt{\pi n_{d}}R_{p}\simeq6$
which implies a mean level spacing in the droplets of $\delta\sim\Delta/N_{e}=6\textrm{K}.$
The distance $r$ between the surfaces of two neighboring droplets
is $r=l_{ip}-2R_{p}\simeq20\,\textrm{nm}.$ The localization length
for inter-droplet tunneling can be obtained from the size of the barrier,
$\xi=\hbar/\sqrt{2mE_{\text{barrier}}}\simeq10\,\textrm{nm},$ which
is of the order of the Bohr radius in GaAs but should be regarded
as a coincidence. For our chosen parameters, $r$ does not exceed
$\xi$ by a large amount.

\section{Magnetic field and density dependence of inter-droplet tunneling
\label{sec:Magnetic-field-dependence}}

Having summarized the physical properties of the electron droplets,
we briefly review their implications for magnetotransport. Unlike
a dirty semiconductor where the electrons are localized at point-like
impurity sites,\cite{Shklovskii1} $l_{ip}$ is comparable with the
droplet diameter. This scenario was studied in Ref. \onlinecite{Glazman}
where it was shown that the resistance between two droplets behaves
as \begin{align}
\frac{\mathcal{R}(B)}{\mathcal{R}(0)}=e^{(B/B_{0})^{2}}\frac{1}{\cosh^{2}(B/B_{1})},\label{eq:raikhglazman}\end{align}
 with \cite{VM1} $B_{0}\sim\phi_{0}/(\pi y_{0}l_{ip}),$ and $B_{1}\sim2\phi_{0}/(\pi l_{ip}^{2}).$
We estimate that the spread of the wavefunction under the barrier
in the direction perpendicular to the tunneling is $y_{0}\sim\sqrt{\xi r}.$
$B_{1}$ is the field below which interference effects are significant.
The magnetoresistance data in Ref.~\onlinecite{Baenninger2} can
be explained with Eq.~(\ref{eq:raikhglazman}) without invoking a
CDW or WC scenario. In particular, $B_{0}^{-2}\simeq A\, r_{ee}^{3}$
which was the primary motivation for suggesting a CDW. The good agreement
with experiment is strong evidence for the existence of electron droplets.
Equation (\ref{eq:raikhglazman}) expresses the inverse of the barrier
transparency between droplets, and at large fields, the transparency
decreases exponentially with increasing $B$. This implies that droplets
become isolated from each other with increasing field, and it is then
natural to expect Coulomb blockade effects will strengthen with magnetic
field, similar to those recently observed in a lattice of quantum
dots. \cite{Dorn}

\section{Magnetic field induced Coulomb blockade \label{sec:Coulomb-blockade}}

We now consider the properties of the magnetic field induced Coulomb
blockade. The charging energy of a single droplet, $E_{c}$, differs
from the bare value $E_{c}^{0}=e^{2}/(8\pi\epsilon_{0}\kappa R_{p})$
(that of a single metallic sphere) if there is a gate voltage $V_{g}$
that couples to the droplet through the gate capacitance $C_{g}$
to give a gate charge $q_{g}=C_{g}V_{g}$, which takes values in the
interval $[0,1/2]$. For non-zero $q_{g}$, $E_{c}(q_{g})=E_{c}^{0}(1-2q_{g})$,
and hence the charging energy takes values between 0 and $E_{c}^{0}\simeq20$
K (for $R_{p}\simeq30$ nm).

In the devices of interest there are many neighboring droplets which
have a depolarizing effect, renormalizing the bare charging energy.
There is charge screening for distances greater than $R_{c}$, and
if $R_{c}\lesssim7R_{p}$ (easily realized in experiment, since $R_{p}\sim0.6R_{c}$),
one can assume the droplet array only has nearest neighbor interactions.
For a hexagonal array, we estimate the effective charging energy as
$E_{c}^{\textrm{eff}}\simeq0.22E_{c}^{0}\simeq4.4$ K, and so expect
strong renormalization of the droplet charging energy.

To calculate the excitation energy of an $N_{e}$-electron droplet,
we need to consider the level separation $\delta$ as well as the
effective charging energy $(1-2q_{g})E_{c}^{\text{eff}}$. Due to
the small size of the droplet, the Fermi energy $E_{F}$ can lie between
levels $\varepsilon_{N_{e}+1}$ and $\varepsilon_{N_{e}}$, corresponding
to $N_{e}+1$ and $N_{e}$ electrons in the droplet respectively.
The energy to add an electron to the droplet is thus: \begin{equation}
E_{\textrm{exc}}^{N_{e}+1,N_{e}}=(1-2q_{g})E_{c}^{\text{eff}}+\textrm{min}\left[\varepsilon_{N_{e}+1}-E_{F},\varepsilon_{N_{e}+1}-\varepsilon_{N_{e}}\right].\label{eq:excitationE}\end{equation}
 The gate voltage tunes both $q_{g}$ and $E_{F}$, so that the first
term oscillates between 0 and $E_{c}^{\text{eff}}$, and the second
between 0 and $\delta$. Equation~(\ref{eq:excitationE}) is for
an isolated droplet, and tunnelling into the droplet will reduce $E_{\textrm{exc}}$.

In a magnetic field, the energies $\varepsilon_{N_{e}}$ are modified
due to both Zeeman splitting and orbital effects. For fields of the
order of a tesla, as is the case in Ref.~\onlinecite{Baenninger2},
the Zeeman splitting $g\mu_{B}B\ll\delta$ can be ignored. Orbital
excitations will generically be affected by a magnetic field, and
when the cyclotron energy $\hbar\omega_{c}=\hbar eB/m\sim\delta$,
one can use Darwin-Fock theory \cite{darwin} to determine the single-particle
energy levels. In our case, for $B=1\,\textrm{T},$ $\hbar\omega_{c}/2\simeq10\,\textrm{K}$
is of the order of $\delta\simeq6\,\textrm{K}.$ However, at our level
of analysis, such an improvement in accuracy does not strongly affect
our results. This is also borne out by experiment, where the position
of the Coulomb blockade peaks/troughs in a resistance versus gate
voltage plot do not shift with field in the magnetic field range considered.\cite{Ilani2}

Treating electron droplets as effectively quantum dots, we now turn
to consider the resistance that arises due to tunneling between droplets
as a function of temperature. The tunneling rate $\Gamma$ between
two droplets or between a droplet and leads is proportional to the
corresponding transmission probability (and hence both droplet spacing
and magnetic field) and the density of states on either side of the
barrier. For sequential tunneling at temperatures $T$ such that $\hbar\Gamma\ll k_{B}T\ll\delta,E_{\text{exc}},$
Coulomb blockade gives rise to the droplet conductance \cite{glazman89,beenakker91}\begin{equation}
G_{CB}=\frac{G_{0}}{\cosh^{2}\left(\frac{E_{\text{exc}}}{2k_{B}T}\right)},\quad G_{0}=\frac{g_{s}e^{2}}{4hk_{B}T}\frac{\Gamma_{l}\Gamma_{r}}{\Gamma_{l}+\Gamma_{r}},\label{eq:GCB}\end{equation}
 where $G_{0}$ is the peak value of the conductance of a droplet
connected through (forward) scattering rates $\Gamma_{l,r}$ with
its left and right neighbors, and $g_{s}$ is the spin degeneracy.
At temperatures lower than $\hbar\Gamma/k_{B},$ $G_{0}$ saturates
to $g_{s}e^{2}/h.$ In Eq.(\ref{eq:GCB}), magnetic field controls
$\Gamma$, whereas density (or $E_{F}$) controls $E_{\textrm{exc}}$.
There are also parallel conductance channels, such as resonant co-tunneling
\cite{ricco84,buttiker86} which is one order higher in the (small)
tunneling probability and significant only at very low temperatures.
For a single droplet (say the $\mathbf{i}^{th}$), the transmission
probability $\mathcal{T}_{\text{cotunn}}$ associated with cotunneling
is (when $k_{B}T\ll\hbar\Gamma$) \begin{align}
\mathcal{T}_{\text{cotunn}}^{(\mathbf{i})} & =\frac{\Gamma_{l}^{(\mathbf{i})}\Gamma_{r}^{(\mathbf{i})}}{2\left(\Gamma_{l}^{(\mathbf{i})}+\Gamma_{r}^{(\mathbf{i})}\right)}\frac{\Gamma^{(\mathbf{i})}}{(E_{\text{exc}}^{(\mathbf{i})}/\hbar)^{2}+(\Gamma^{(\mathbf{i})}/2)^{2}},\label{eq:T-cotunn}\end{align}
 where the total decay width $\Gamma^{(\mathbf{i})}$ includes inelastic
as well as the elastic contributions, $\Gamma_{l,r}^{(\mathbf{i})}.$
As inelastic processes are usually present, $\Gamma^{(\mathbf{i})}>\Gamma_{l}^{(\mathbf{i})}+\Gamma_{r}^{(\mathbf{i})}.$

For a string of $\mathcal{N}$ droplets, the total cotunneling transmission
$\mathcal{T}$ is the product of the cotunneling transmissions, $\mathcal{T}=\prod_{i=1}^{\mathcal{N}}\mathcal{T}_{\text{cotunn}}^{(\mathbf{i})}$
of individual droplets in the string, and hence the cotunneling conductance
is $G_{\text{cotunn}}=g_{s}(e^{2}/h)\mathcal{T}.$ This contribution
is thus small unless there are uniformly spaced identical droplets,
in which case there can be a contribution at resonance. \cite{buttiker86}

\begin{figure}
\includegraphics[width=6cm,keepaspectratio,angle=270]{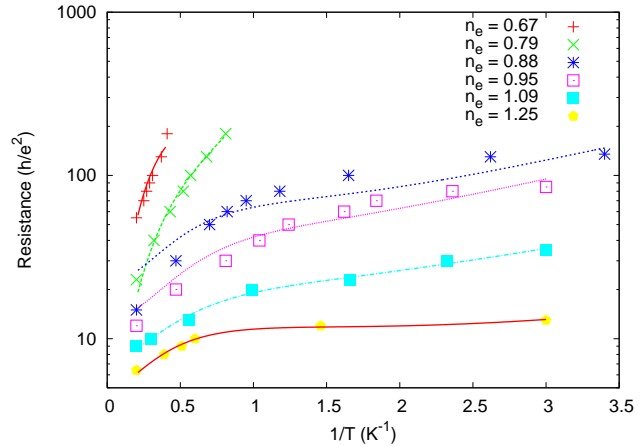}

\caption{\label{fig:CBconductance} (Color online) Temperature dependence
of resistance data from Ref.~\onlinecite{Baenninger2} fitted to
Eq.~(\ref{eq:singledropletR}) with an additional series resistance.
The fits yield an excitation energy $E_{\text{exc}}\sim1\,\textrm{K}\sim\delta_{\text{sample}}.$
The electron densities, $n_{e}$, are shown in units of $10^{10}\,\textrm{cm}^{-2}.$}
\end{figure}

The relative importance of cotunneling and activated conduction changes
as the droplet separation, $l_{ip}-2r$, decreases towards $\xi$,
implying that $\hbar\Gamma$ approaches $E_{\text{exc}}.$ This allows
cotunneling to contribute to conductance at low temperatures. Assuming
both contributions acting in parallel we estimate the droplet resistance
as

\begin{equation}
\mathcal{R}\simeq\frac{\cosh^{2}\left(\frac{E_{\text{exc}}}{2k_{B}T}\right)}{G_{0}+G_{\text{cotunn}}\cosh^{2}\left(\frac{E_{\text{exc}}}{2k_{B}T}\right)}.\label{eq:singledropletR}\end{equation}
 Equation~(\ref{eq:singledropletR}) is not valid in the high or
very low temperature limits since it implies $\mathcal{R}\rightarrow0$
as $1/T\rightarrow0$, and does not include the fact that $G_{0}$
saturates as $T\rightarrow0$. We use Eq.~(\ref{eq:singledropletR})
to fit resistance versus T data from Ref.~\onlinecite{Baenninger2},
and in so-doing additionally assume a series resistance $R_{0}$,
so that the total resistance $R=R_{0}+\mathcal{R}$. We also assume
a phenomenological form for $G_{\textrm{cotunn}}=G_{\textrm{cotunn}}(T=0)(1+cT+aT^{2})$,
with $a$ and $c$ non-negative fitting parameters. Physically, the
quadratic-$T$ dependence comes from inelastic cotunneling and the
linear-$T$ dependence arises from the linear suppression of the tunneling
density of states due to the Anderson orthogonality catastrophe when
the droplet is coupled asymmetrically to the leads, \cite{furusaki95}
which is the generic situation. We find excellent agreement with experiment
at almost all densities, as is evident in Fig.~\ref{fig:CBconductance}
and we extract $E_{\textrm{exc}}\simeq1$ K at most values of $n_{e}$.
We note that extrapolation of the resistance in the Coulomb blockade
regime to $B=0$ leads to a resistance that is always greater than
$h/2e^{2}$, implying that there is, effectively, never more than
one conductance channel open for transport, in agreement with the
success of Eq.~(\ref{eq:singledropletR}) in fitting the data.

\begin{figure}
\includegraphics[width=6cm,keepaspectratio,angle=270]{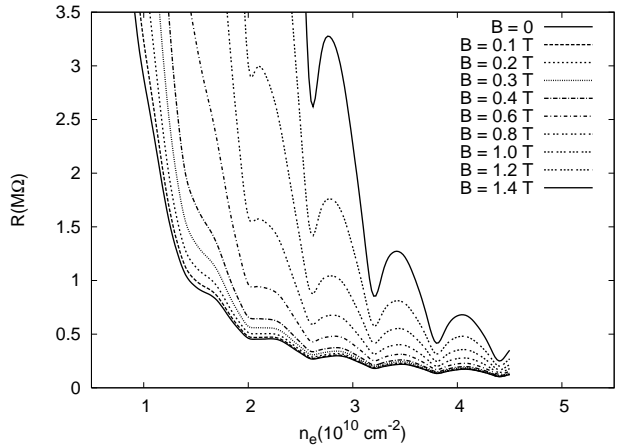}

\caption{\label{fig:schematic} Plot showing magnetic field induced Coulomb
blockade calculated using Eq.~(\ref{eq:singledropletR}). For small
values of the magnetic field, the inter-droplet tunneling is strong
enough to wash out Coulomb blockade effects. Large magnetic fields
reduce the inter-droplet tunneling by shrinking the localization length,
leading to an enhanced visibility of Coulomb blockade oscillations. }
\end{figure}

As $V_{g}$ is varied, the droplet energy levels cross $E_{F}$, and
since $E_{\text{exc}}$ also depends on $V_{g}$ {[}see Eq.~(\ref{eq:excitationE}){]},
the excitation energy for a droplet will vary between 0 and $E_{c}^{\text{eff}}+\delta.$
If there are many droplets, which we believe is the case here, we
assume that the excitation energies are uniformly distributed in the
interval $[0,E_{c}^{\text{eff}}+\delta].$ In Fig.~\ref{fig:schematic}
we plot resistance as a function of $n_{e}$ for illustrative purposes.
We used Eqs.~(\ref{eq:GCB}), (\ref{eq:T-cotunn}) and (\ref{eq:singledropletR}),
along with the schematic form $E_{\textrm{exc}}=\frac{E_{0}}{2}\left[1+f\cos\left(\frac{2\pi(n_{e}-n_{0})}{\Delta n}\right)\right]$,
where $0\lesssim f\lesssim1$ (we choose $f=0.6$ here), $E_{0}=1$K,
$T=1$K, $\Delta n=0.6\times10^{10}\,\textrm{cm}^{-2}$, and $n_{0}=0.5\times10^{10}\,\textrm{cm}^{-2}$.
We assume that $\Gamma_{l}=\Gamma_{r}=\Gamma$ and the $n_{e}$ and
$B$ dependence of $\Gamma$ is determined by Eq.~(\ref{eq:raikhglazman}).
The behaviour seen in Fig.~\ref{fig:schematic} can be deduced very
easily by looking at limits of Eq.~(\ref{eq:singledropletR}). In
the limit that $G_{0}\ll G_{\textrm{cotunn}}$, $\mathcal{R}\sim\frac{1}{G_{\textrm{cotunn}}}$,
whereas when $G_{0}\gg G_{\textrm{cotunn}}$, $\mathcal{R}\sim\frac{1}{G_{0}}\cosh^{2}\left(\frac{E_{\textrm{exc}}}{2k_{B}T}\right)$.
We expect that both $G_{\textrm{cotunn}}$ and $G_{0}$ will decrease
with $n_{e}$, but $G_{0}$ will decrease faster since $\Gamma^{({\rm i})}$
includes inelastic as well as elastic contributions, which are likely
to be less density dependent. This would imply a crossover from oscillations
to increasing resistance with decreasing density, exactly as shown
in Fig.~\ref{fig:schematic}.

The temperature dependence of thermally activated transport depends
on the size of the sample. For a sample large enough to contain many
droplets there will be some with arbitrarily low activation energy,
in which case transport will proceed via variable range hopping (VRH),
\cite{Mott,AHL,ES} and we do not expect to see Coulomb blockade oscillations.\cite{fowler}
The crossover length for Arrhenius to Mott VRH behavior may be estimated
as follows. For Mott VRH in 2$d$, the characteristic hopping distance
$D_{\text{Mott}}$ is given by $D_{\text{Mott}}=\left(\xi/\pi\nu k_{B}T\right)^{1/3},$
where $\nu$ is the density of available localized states per unit
area: \begin{eqnarray}
\nu & \approx\frac{1}{\pi(l_{ip}/2)^{2}(\delta+E_{c}^{\text{eff}})}.\label{eq:density_states}\end{eqnarray}
 This gives $D_{\text{Mott}}\simeq60\, T^{-1/3}\,\textrm{nm\, K}^{-1/3}$,
and at $T=50\,\textrm{mK}$, $D_{\text{Mott}}\simeq160\,\textrm{nm}.$
In Ref.~\onlinecite{Baenninger2}, an Arrhenius law in temperature
is observed, implying that the active device area should be less than
$D_{\text{Mott}}^{2}.$ The total number of droplets $N_{p}$ in an
area $\pi D_{\text{Mott}}^{2}$ is \begin{eqnarray}
N_{p} & =\frac{4D_{\text{Mott}}^{2}}{l_{ip}^{2}}=\left(\frac{2\xi(\delta+E_{c}^{\text{eff}})}{l_{ip}k_{B}T}\right)^{2/3},\label{eq:NP}\end{eqnarray}
 which is about $15$ for the typical parameters we assume. Some low
density samples do show VRH behavior,\cite{Ghosh02} and hence we
assume that the effective size of the sample is of the order of $D_{\text{Mott}},$
so the mean level separation in the sample is approximately $\delta_{\text{sample}}\approx\frac{(E_{c}^{\text{eff}}+\delta)}{N_{p}},$
which is about $0.75\,\textrm{K}$. The separation of the magnetoresistance
peak from a trough corresponds to an energy scale of $\simeq1\textrm{K},$
in good agreement with both the activation energy deduced from experiment,
and with $\delta_{\text{sample}}$.

As $V_{g}$ is varied, excited states in different droplets successively
come into resonance; these resonances are associated with the minima
of resistance. In between the minima, if the sample is small, no droplet
in the system is in resonance with $E_{F}$ and the resistance will
show a maximum. The observability of resistance oscillations therefore
crucially depends on the samples being small. This seems to be borne
out in experiment. \cite{Baenninger2}

At large electron densities, the oscillations will be less visible
for two reasons; firstly, as $n_{e}$ increases, $l_{ip}$ decreases,
as does $E_{c}^{\textrm{eff}}$, which reduces $\delta_{\text{sample}}$.
This reduces the contrast between resonant and Coulomb blockaded states.
Secondly, the inter-droplet tunneling distance, $D\propto1/\sqrt{n_{e}}$,
decreases \cite{VM1} and hence $\Gamma$ increases. Ultimately, when
the localization condition, $D/\xi\gtrsim1,$ cannot be satisfied,
the system becomes well-conducting and no Coulomb blockade oscillations
are possible.

Increasing magnetic field reduces $\Gamma$, with relatively little
effect on excitation energies which improves the visibility of the
Coulomb blockade. In experiment, the resistance oscillations tend
to appear above a threshold field, which we estimate to be where the
magnetoresistance switches from negative to positive, in the vicinity
of $B_{1}\propto n_{e},$ where $B_{1}$ is defined below Eq.~(\ref{eq:raikhglazman}).

\section{Discussion \label{sec:Discussion}}

In some respects, the issues discussed here are similar to observations
in \emph{one dimensional} wires of variations in the conductance periodic
in $n_{e}$. \cite{SThomas} These were initially described in terms
of a CDW, whereas later investigations appear to have convincingly
demonstrated that the oscillations are due to Coulomb blockade effects.
\cite{comment,Staring} It was found that the Coulomb blockade effects
strengthened as a magnetic field was applied, consistent with the
picture proposed here.\cite{Staring} However, the Coulomb blockade
did not rely on magnetic field for visibility.

We did not discuss how non-linear screening is affected by the presence
of a magnetic field. The screening of 2D electrons in a disordered
potential in a magnetic field was discussed in Refs.~\onlinecite{Efros1,EPB},
focusing on the regime where disorder is not too strong. We note that
experiment appears to provide some of the solution. Measurements of
localized states in the quantum Hall regime \cite{Ilani2} that support
a dot-like picture at low densities find the local electronic compressibility
to be essentially independent of magnetic field. We also ignored possible
correlation of donor charges in the dopant layer -- including these
worsened the agreement with experiment.\cite{VM1} However, donor
correlations are likely to be relevant in some cases.\cite{AriNP}

In summary, we use the picture of electron droplets to establish that
for low density 2DEGs in disordered delta doped heterostructures there
can be a magnetic field induced Coulomb blockade. We provide evidence
for this picture by using a model for resistance as a function of
temperature based on the idea of electron droplets acting like quantum
dots to successfully fit experimental data. The ideas we present here
may have wider applicability in inhomogeneous strongly correlated
electron systems.

We thank D. Khmelnitskii, N. Cooper, A. Ghosh and M. Baenninger for
useful discussions, and A. Ghosh and M. Baenninger for sharing unpublished
data. V.T. thanks TIFR and DST Ramanujan Fellowship {[}Sanction No.
100/IFD/154/2007-08{]}, and M.P.K. thanks NSERC and an SFU President's
Research Grant for support. We also acknowledge the support of Trinity
College, Cambridge where this work was begun.


\begin{thebibliography}{10}
\bibitem{Baenninger2}M. Baenninger, A. Ghosh, M. Pepper, H. E. Beere, I. Farrer, P. Atkinson,
and D. A. Ritchie, Phys. Rev. B \textbf{72}, 241311(R) (2005).
\bibitem{Ghosh2}A. Ghosh, M. Pepper, H. E. Beere, and D. A. Ritchie, Phys. Rev. B
\textbf{70}, 233309 (2004).
\bibitem{Ghosh3}A. Ghosh, C. J. B. Ford, M. Pepper, H. E. Beere, and D. A. Ritchie,
Phys. Rev. Lett. \textbf{92}, 116601 (2004).
\bibitem{Ilani2}S. Ilani, J. Martin, E. Teitelbaum, J. H. Smet, D. Mahalu, V. Umansky,
and A. Yacoby, Nature \textbf{427}, 328 (2004).
\bibitem{Dorn}A. Dorn, T. Ihn, K. Ensslin, W. Wegscheider, and M. Bichler, Phys.
Rev. B \textbf{70}, 205306 (2004).
\bibitem{Baenninger3}M. Baenninger, A. Ghosh, M. Pepper, H. E. Beere, I. Farrer, P. Atkinson,
and D. A. Ritchie, electronic preprint arXiv:0707.3543.
\bibitem{Tanatar}B. Tanatar and D. M. Ceperley, Phys. Rev. B \textbf{39}, 5005 (1989);
S. T. Chui and B. Tanatar, Phys. Rev. Lett. \textbf{74}, 458 (1995).
\bibitem{Suris}V. A. Gergel' and R. A. Suris, Zh. Eksp. Teor. Fiz. \textbf{75}, 191
(1978). {[}Sov. Phys. JETP \textbf{48}, 95 (1978){]}
\bibitem{droplets}J. A. Nixon and J. H. Davies, Phys. Rev. B \textbf{41}, 7929 (1990);
J. Shi and X. C. Xie, Phys. Rev. Lett. \textbf{88}, 086401 (2002);
M. M. Fogler, Phys. Rev. B \textbf{69}, 121409(R) (2004); \textit{ibid}.
\textbf{69}, 245321 (2004); \textit{ibid.} \textbf{70}, 129902(E)
(2004); J. Wiebe, Chr. Meyer, J. Klijn, M. Morgenstern, and R. Wiesendanger,
\textit{ibid}. \textbf{68}, 041402 (2003).
\bibitem{VM1}V. Tripathi and M. P. Kennett, Phys. Rev. B \textbf{74}, 195334 (2006).
\bibitem{hall-droplets}S. Ilani, A. Yacoby, D. Mahalu, H. Shtrikman, Science \textbf{292},
1354 (2001); J. Martin, S. Ilani, B. Verdene, J. Smet, V. Umansky,
D. Mahalu, D. Schuh, G. Abstreiter, and A. Yacoby, \textit{ibid.}
\textbf{305}, 980 (2004).
\bibitem{kivelson}B. Spivak and S. A. Kivelson, \textit{ibid}. \textbf{70}, 155114 (2004);
B. Spivak ans S. Kivelson, J. de Physique IV \textbf{131}, 255 (2005);
R. Jamei, S. Kivelson, and B. Spivak, Phys. Rev. Lett. \textbf{94},
056805 (2005).
\bibitem{Davis}S. H. Pan, J. P. O'Neal, R. L. Badzey, C. Chamon, H. Ding, J. R. Engelbrecht,
Z. Wang, H. Eisaki, S. Uchida, A. K. Gupta, K.-W. Ng, E. W. Hudson,
K. M. Lang, and J. C. Davis, Nature \textbf{413}, 282 (2001); K. M.
Lang, V. Madhavan, J. E. Hoffman, E. W. Hudson, H. Eisaki, S. Uchida,
and J. C. Davis, Nature \textbf{415}, 412 (2002); N. Mathur and P.
B. Littlewood, Phys. Today \textbf{56}, 25 (2003); S. Nakatsuji, V.
Dobrosavljevic, D. Tanaskovic, M. Minakata, H. Fukazawa, and Y. Maeno,
Phys. Rev. Lett. \textbf{93}, 146401 (2004).
\bibitem{Mcelroy}K. McElroy, Jinho Lee, J. A. Slezak, D.-H. Lee, H. Eisaki, S. Uchida,
and J. C. Davis, Science \textbf{309}, 1048 (2005).
\bibitem{Shklovskii1}B. I. Shklovskii and A. L. Efros, Zh. Eksp. Teor. Fiz. \textbf{84},
811 (1983). {[}Sov. Phys. JETP \textbf{57}, 470 (1983){]}; B. I. Shklovskii,
Fiz. Tekh. Poluprovdn \textbf{17}, 2055 (1983). {[}Sov. Phys. Semicond.
\textbf{17}, 1311 (1983){]}.
\bibitem{Glazman}M. E. Raikh and L. I. Glazman, Phys. Rev. Lett. \textbf{75}, 128 (1995).
\bibitem{darwin}C. G. Darwin, Proc. Cambridge Phil. Soc. \textbf{27}, 86 (1930); V.
Fock, Z. Phys. \textbf{47}, 446 (1928).
\bibitem{glazman89}L. I. Glazman and R. I. Shekhter, J. Phys. Cond. Mat. \textbf{1},
L5811 (1989).
\bibitem{beenakker91}C. W. J. Beenakker, Phys. Rev. B \textbf{44}, 1646 (1991).
\bibitem{ricco84}B. Ricco and M. Ya. Azbel, Phys. Rev. B \textbf{29}, 1970 (1984);
A. D. Stone and P. A. Lee, Phys. Rev. Lett. \textbf{54}, 1196 (1985).
\bibitem{buttiker86}M. Büttiker, Phys. Rev. B \textbf{33}, 3020 (1986).
\bibitem{furusaki95}A. Furusaki and K. A. Matveev, Phys. Rev. B \textbf{52}, 16676 (1995).
\bibitem{Mott}N. F. Mott, J. Non-Cryst. Solids \textbf{1}, 1 (1968).
\bibitem{AHL}V. Ambegoakar, B. I. Halperin, and J. S. Langer, Phys. Rev. B \textbf{4},
2612 (1971).
\bibitem{ES}A. L. Efros and B. I. Shklovskii, J. Phys. C \textbf{8}, L49 (1975).
\bibitem{fowler}A. B. Fowler, Proc. Natl. Acad. Sci. USA \textbf{84}, 4701 (1987).
\bibitem{Ghosh02}A. Ghosh, M. Pepper, D. A. Ritchie, E. H. Linfield, R. H. Harrell,
H. E. Beere, and G. A. C. Jones, Phys. Stat. Sol. B \textbf{230},
211 (2002).
\bibitem{SThomas}J. H. F. Scott-Thomas, S. B. Field, M. A. Kastner, H. I. Smith, and
D. A. Antoniadis, Phys. Rev. Lett. \textbf{62}, 583 (1989); M. A.
Kastner, S. B. Field, U. Meirav, J. H. F. Scott-Thomas, D. A. Antoniadis,
and H. I. Smith, \textit{ibid} \textbf{63}, 1894 (1989); U. Meirav,
M. A. Kastner, M. Heiblum, and S. J. Wind, Phys. Rev. B \textbf{40},
5871(R) (1989).
\bibitem{comment}H. van Houten and C. W. J. Beenakker, Phys. Rev. Lett. \textbf{63},
1893 (1989).
\bibitem{Staring}A. A. M. Staring, H. van Houten, C. W. J. Beenakker, and C. T. Foxon,
Phys. Rev. B \textbf{45}, 9222 (1992).
\bibitem{Efros1}A. L. Efros, Solid State Commun. \textbf{67}, 1019 (1988).
\bibitem{EPB}A. L. Efros, F. G. Pikus, and V. G. Burnett, Phys. Rev. B \textbf{47},
2233 (1993).
\bibitem{AriNP}C. Siegert, A. Ghosh, M. Pepper, I. Farrer, and D. A. Ritchie, Nature
Phys. \textbf{3}, 315 (2007). 
\end{thebibliography}
\end{document}